\documentclass[12pt]{iopart}

\usepackage{iopams}
\usepackage{epsfig}
\begin{document}

\title[]{Angular hadron correlations probing the early medium evolution}

\author{Thorsten Renk and Kari J.~Eskola}

\address{Department of Physics, PO Box 35 FIN-40014 University of Jyv\"askyl\"a, Finland and}
\address{Helsinki Institut of Physics, PO Box 64 FIN-00014, University of Helsinki, Finland}

\ead{trenk@phys.jyu.fi,kari.eskola@phys.jyu.fi}

\begin{abstract}
Hard processes are a well calibrated probe to study heavy-ion collisions. However, the information to be gained from the nuclear suppression factor $R_{AA}$ is limited, hene one has to study more differential observables to do medium tomography. The angular correlations of hadrons associated with a hard trigger appear suitable as they  show a rich pattern when going from low $p_T$ to high $p_T$. Of prime interest is the fate of away side partons with an in-medium pathlength $O$(several fm). At high $p_T$ the correlations become dominated by the punchtrough of the away side parton with subsequent fragmentation. We discuss what information about the medium density can be gained from the data. 

\end{abstract}


\section{Introduction}

Considerable effort is made in the study of the energy loss of hard partons in the hot and dense medium created in ultrarelativistic heavy-ion collisions. In particular medium tomography (i.e. gaining information about the medium density evolution) by studying the nuclear  suppression factor $R_{AA}$, the observed transverse momentum spectrum of hard hadrons divided by the scaled expectation from p-p collisions \cite{Jet1}. However, as argued in 
\cite{Gamma-Tomography}, $R_{AA}$ exhibits very limited sensitivity to the energy loss 
mechanism or properties of the medium beyond the fact that the quenching is substantial.

If one makes the ansatz that the local quenching power $\hat{q}$ of the medium can be written as a function of the local energy density $\epsilon$ as \cite{Baier}
$\hat{q}(\tau, \eta_s, r,\phi) = K \cdot 2 \cdot [\epsilon(\tau, \eta_s, r, \phi)]^{3/4}$
where $\tau$ is evolution proper time, $\eta_s$ spacetime rapidity and $r$ radius of the medium, adjusting the scale parameter $K$ parametrizes both the coupling strength of the parton to the medium and uncertainties in the determination of the medium evolution (see \cite{Correlations}). In \cite{ReactionPlane} it was shown that there is about 50\% uncertainty in the extraction of the medium quenching power (given by $K$) from $R_{AA}$ even for model evolutions which describe bulk soft matter properties; a 2+1d hydrodynamical evolution \cite{Hydro}, a 3+1d hydrodynamical evolution \cite{Hydro3d} and a parametrized evolution \cite{Parametrized}. Greater sensitivity to the medium density evolution can hence only be gained in more differential probes such as $\gamma$-hadron correlations \cite{Gamma-Tomography}, $R_{AA}$ for non-central collisions \cite{ReactionPlane} or back-to-back hadron correlations \cite{Correlations}, here we focus on the latter.

\section{The model description}

We make a Monte Carlo (MC) simulation of hard back-to-back processes in a medium in order to calculate the induced hard hadronic correlation pattern. The model is described in detail in \cite{Correlations}. There are three main elements: 1) the primary hard pQCD process 2) the description of the soft medium and 3) the energy loss from hard to soft degrees of freedom coupling 1) and 2). 

The energy loss for a given parton path is described in a probabilistic language.
In order to determine the probability  $P(\Delta E,  E)_{path}$ for a 
hard parton with energy $E$ to lose the energy $\Delta E$ while traversing the medium on its 
trajectory, we make use of a scaling law \cite{JetScaling} which allows to relate the dynamical 
scenario to a static equivalent one for each trajectory $\xi(\tau)$ by calculating 
 
\begin{equation}
\label{E-omega}
\omega_c({\bf r_0}, \phi) = \int_0^\infty d \xi \xi \hat{q}(\xi)
\quad {\rm and} \quad
\langle\hat{q}L\rangle ({\bf r_0}, \phi) = \int_0^\infty d \xi \hat{q}(\xi)
\end{equation} 
 
as a function of the jet production vertex ${\bf r_0}$ and its angular orientation $\phi$. We set 
$\hat{q} \equiv 0$ whenever the decoupling temperature of the medium $T = T_F$ is reached.
Using the numerical results of \cite{QuenchingWeights}, we obtain $P(\Delta E)_{path}$ 
for $\omega_c$ and $R=2\omega_c^2/\langle\hat{q}L\rangle$ 
as a function of jet production vertex and the angle $\phi$.

From this, the probability distribution of energy loss reflected in single hadron observables (such as $R_{AA}$) can be computed by averaging over all possible production points determined by the nuclear overlap as

\begin{equation}
\label{E-P_TAA}
\langle P(\Delta E)\rangle_{T_{AA}} =\frac{1}{2\pi} \int_0^{2\pi}  
 d\phi 
\int_{-\infty}^{\infty}  dx_0 
\int_{-\infty}^{\infty} dy_0 P(x_0,y_0)  
P(\Delta E)_{path}.
\end{equation}

The averaged energy loss probability underlying dihadron correlations involves a conditional probability of observing a high $p_T$ trigger hadron, hence it is more complicated and must be computed in the full MC simulation. We do the simulation generating hard pQCD back-to-back events inside the evolving bulk matter model, using Eq.~(\ref{E-omega}) to determine $P(\Delta E)_{path}$ for each parton, followed by leading and next to leading fragmentation once the parton emerges from the medium \cite{Correlations}. In all simulations, the scale parameter $K$ is adjusted such that the measured $R_{AA}$ is described well, thus we are only interested in information beyond what can be extracted from single hadron observables.

We test six different spacetime evolution scenarios: 1) a 2+1d Bjorken hydrodynamics simulation ('hydrodynamics') 2) a 2+1 d Bjorken hydrodynamics evolution in which quenching occurs only in the quark-gluon plasma phase ('black core'), 3) a parametrized non-Bjorken expansion with box density 4) a parametrized non-Bjorken expansion with nuclear profile $T_A$ density 5) a Bjorken expansion with box density and 6) a Bjorken expansion with $T_A$ density \cite{Correlations}. In this way, we represent a selection of different longitudinal and transverse dynamics.

\section{Results}

\begin{figure*}[htb]
\epsfig{file=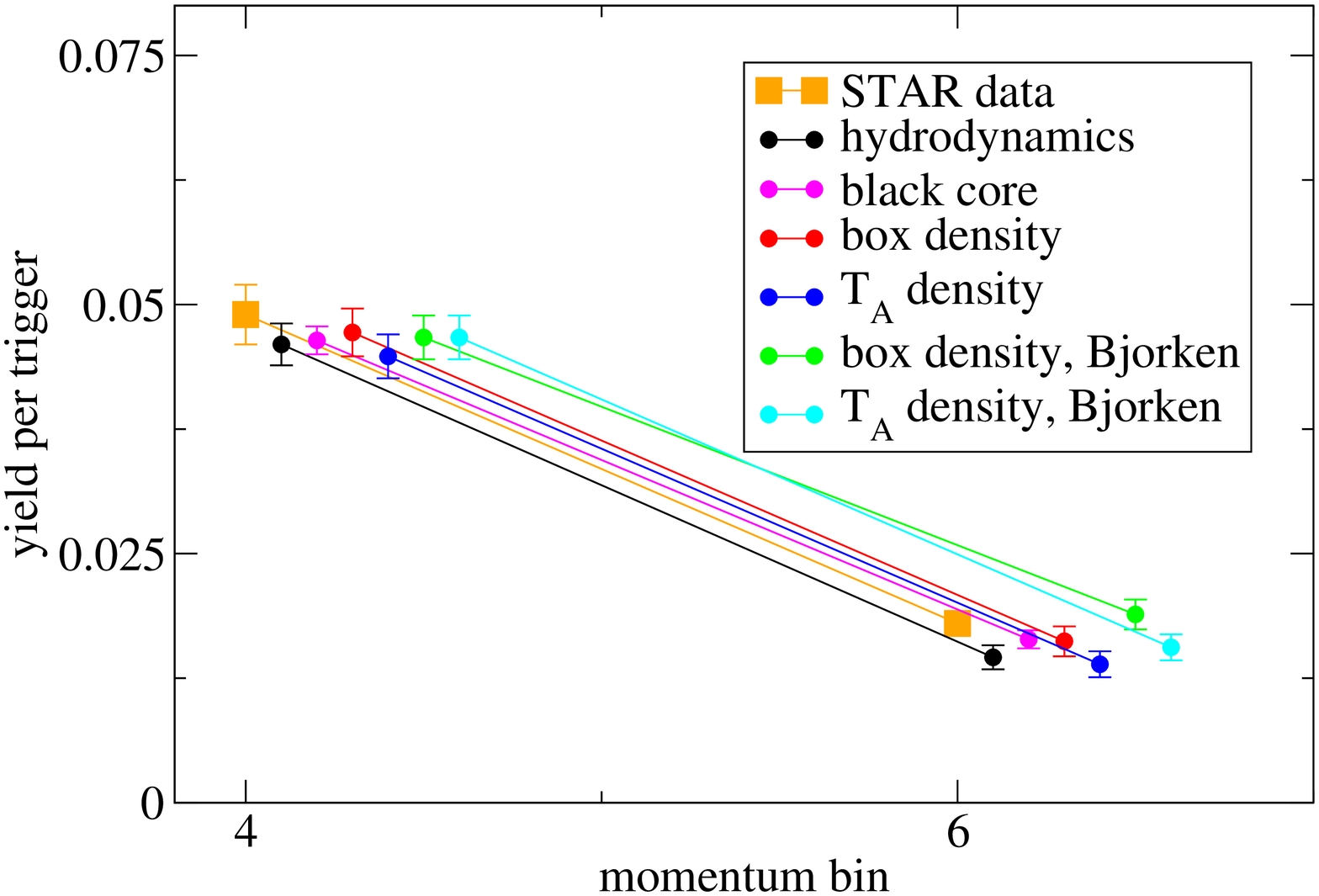, width=7.6cm}\epsfig{file=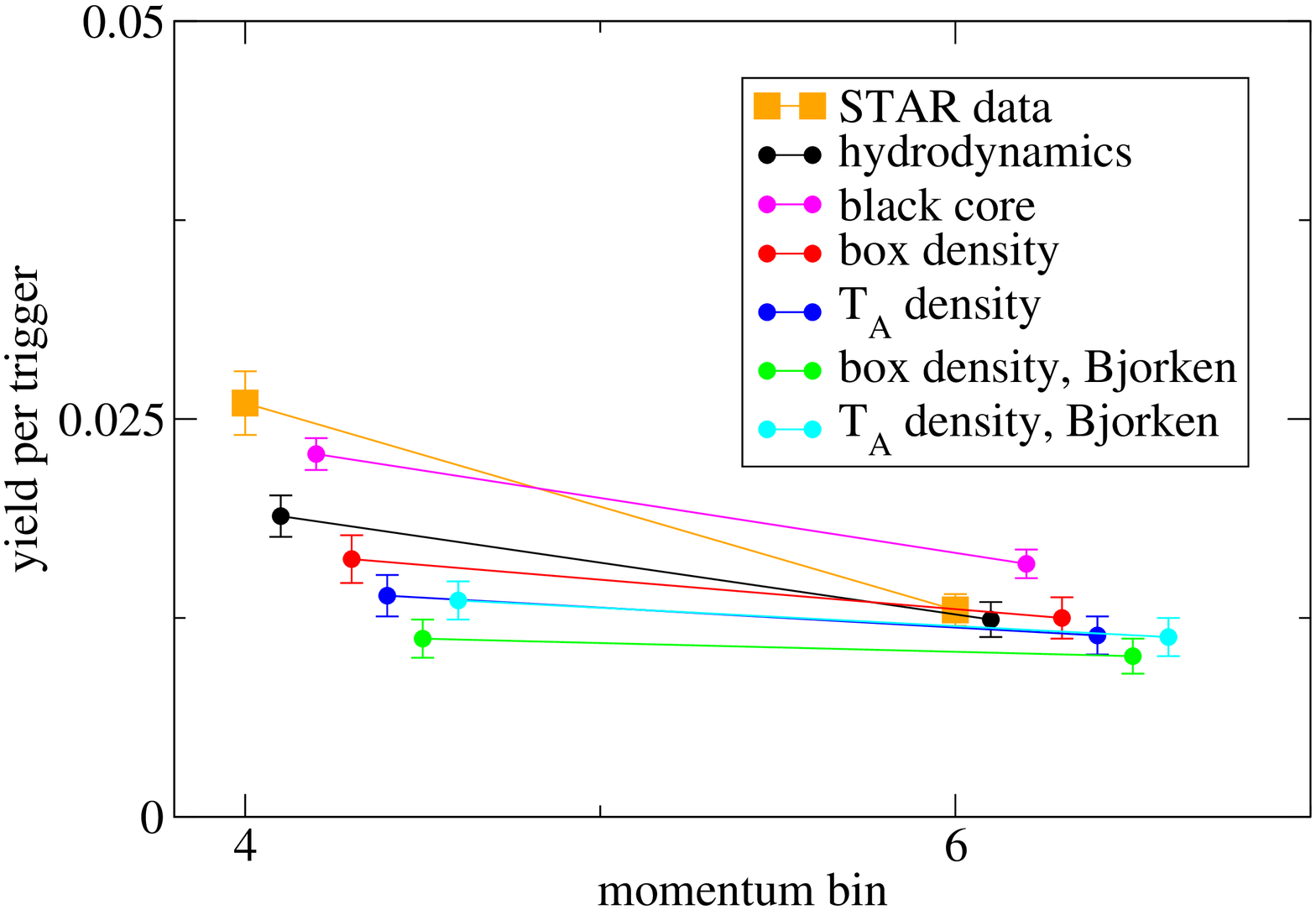, width=7.6cm}
\caption{\label{F-ypt8}Yield per trigger on the near side (left) and away side (right)  
of hadrons in the 4-6 GeV and 6+ GeV momentum bin associated with a trigger with 8 $<  
p_T < $ 15 GeV for different evolution models compared with STAR data  
\cite{Dijets1,Dijets2}. Individual data points are spread artificially along the $x$ axis  
for clarity.}
\end{figure*}

We compare the computed yield of associated hadrons as a function of momentum bin per trigger in the range $8< p_T < 15$ GeV on near and away side with the STAR data \cite{Dijets1,Dijets2} in Fig.~\ref{F-ypt8}. The near side is described well and there is broad agreement across all different medium models. The reason is that in all cases $P(\Delta E)_{T_{AA}}$ is dominated by either transmission without energy loss or complete absorption, the kinematic window to observe shifts in parton energy is very narrow. However, the probability distribution of experiencing a shift $\Delta E$ in energy is where tomographical information resides, thus, all scenarios appear similar for a small difference between trigger and associate $p_T$.

Since the away side yield is determined by a different (conditional) probability $P(\Delta E)_{trigger}$ the models exhibit more pronounced differences. All models fail in the momentum window between 4 and 6 GeV to describe the data. This is presumably due to the fact that pQCD and fragmentation is not the dominant mechanism of hadron production in this momentum regime. Within errors, all scenarios except the 'black core' one agree with the data above 6 GeV associate momentum.

\begin{figure*}[htb]
\epsfig{file=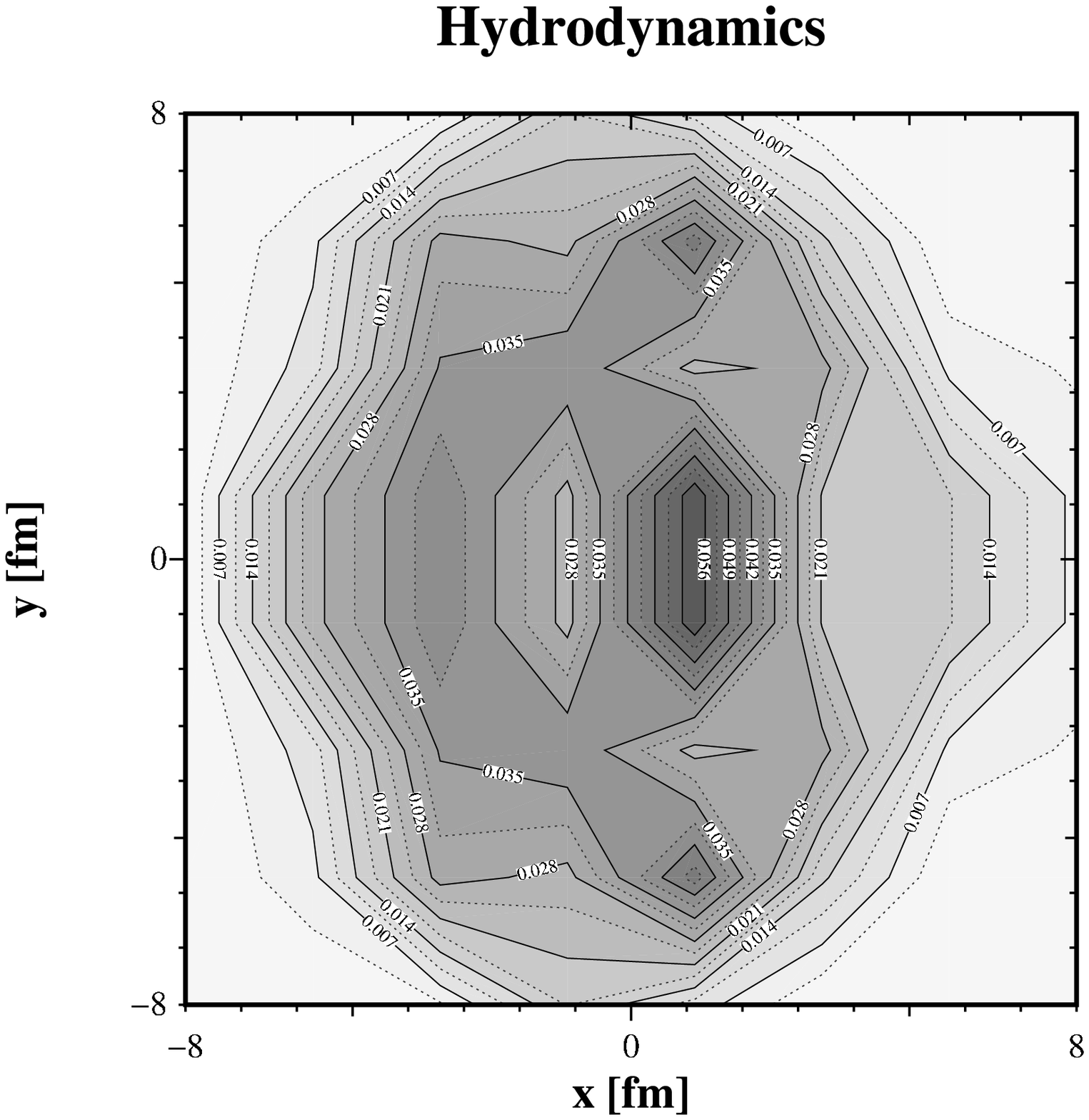, width=7.5cm}\epsfig{file=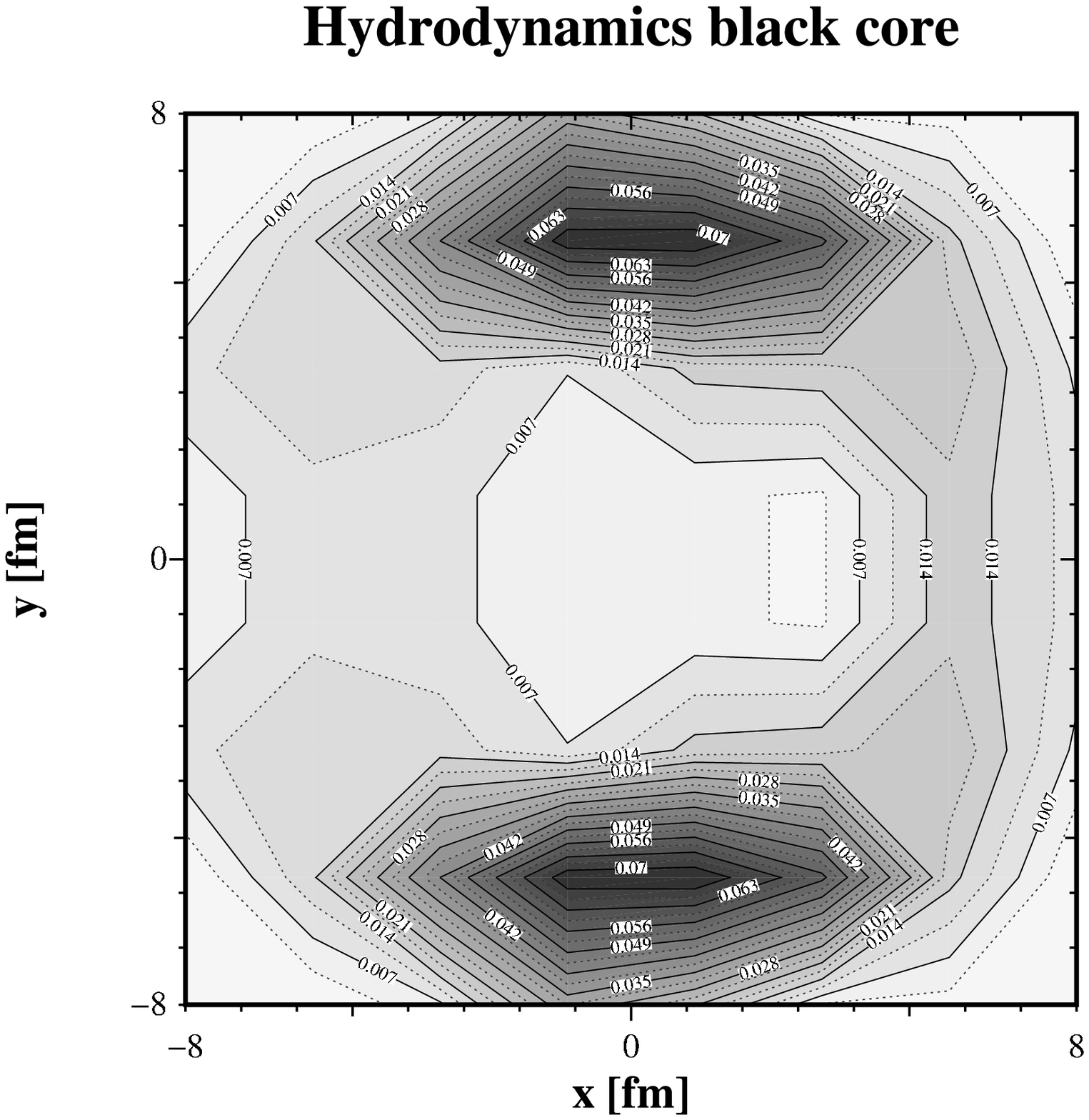, width=7.5cm}
\vspace*{-2.5cm}
\caption{\label{F-Geo}Probability density for finding a vertex at $(x,y)$ leading to a 
triggered event with 8 $< p_T <$ 15 GeV and an away side hadron with 4 $< p_T < 
6$ GeV for different spacetime evolution scenarios. The near side hadron 
propagates to the $-x$ direction.}
\end{figure*}

The black core scenario is characterized by a very high quenching power (48 GeV$^2$/fm at $\tau$ = 1 fm/c) in the center and a large halo out of which partons can escape unhindered. On the other hand, the hydrodynamics scenario shows less quenching power (11 GeV$^2$/fm at $\tau$ = 1 fm/c) but has a larger extension. We show the origin of dihadrons in each model in Fig.~\ref{F-Geo}. It is evident that dihadron production in the black core scenario is peripheral while in the hydrodynamics scenario it occurs in the whole medium. From the fact that only one of these scenarios ('hydrodynamics') describes the data, we can infer that the produced matter is not characterized by an opaque core and a dilute halo and that dihadron emission is not peripheral but probes indeed the central region of the medium.
While this is some promising start, a larger lever arm in trigger $p_T$ would be needed to be sensitive to shifts in parton energy instead of just absorption and to get more tomographical information \cite{Correlations}. 
Recently, elastic energy loss has been suggested as an additional ingredient to describe the nuclear suppression \cite{Elastic}. To the degree that the magnitude of  this contribution is parametrically given by $\Delta E \sim \int_0^\infty d \xi \hat{q}(\xi)$, these results place tight limits on the importance of collisional energy loss: Since the integral has virtually no contribution from later times (long paths), elastic energy loss cannot resolve spatial structures larger than $\sim 2-3$ fm whereas the observed difference between near and away side yield relies on the fact that pathlength differences of the order of 5 fm or more are resolved.

\end{document}